\documentclass{SciPost}

\hypersetup{
    colorlinks,
    linkcolor={red!50!black},
    citecolor={blue!50!black},
    urlcolor={blue!80!black}
}

\usepackage[bitstream-charter]{mathdesign}
\DeclareSymbolFont{usualmathcal}{OMS}{cmsy}{m}{n}
\DeclareSymbolFontAlphabet{\mathcal}{usualmathcal}

\fancypagestyle{SPstyle}{
\fancyhf{}
\lhead{\colorbox{scipostblue}{\bf \color{white} ~SciPost Physics }}
\rhead{{\bf \color{scipostdeepblue} ~Submission }}

\fancyfoot[C]{\textbf{\thepage}}
}

\newcommand{\be}{\begin{equation}}
\newcommand{\ee}{\end{equation}}
\newcommand{\bea}{\begin{eqnarray}}
\newcommand{\eea}{\end{eqnarray}}

\newcommand{\mc}{\mathcal}

\begin{document}

\pagestyle{SPstyle}

\begin{center}{\Large \textbf{\color{scipostdeepblue}{
%%%%%%%%%% TODO: Write your article's title here
 Energy Partitioning of Wave Packets in One-Dimensional Systems \\
%%%%%%%%%% END TODO: TITLE
}}}\end{center}

\begin{center}\textbf{
Fl\'avia B. Ramos\textsuperscript{1},
Imke Schneider\textsuperscript{2},
Sebastian Eggert\textsuperscript{2} and
Rodrigo G. Pereira\textsuperscript{1}
}\end{center}

\begin{center}
{\bf 1} International Institute of Physics, Natal, RN, 59078-970, Brazil
\\
{\bf 2} Physics Department and Research Center OPTIMAS, Rheinland Pf\"alzische Technische Universit\"at, 67663 Kaiserslautern, Germany
\end{center}

\section*{\color{scipostdeepblue}{Abstract}}
\textbf{\boldmath{%
We investigate the non-equilibrium dynamics of wave packets in a one-dimensional critical fermionic system and analyze the resulting partitioning of energy between emergent excitations. Starting from a Gaussian wave packet injected on top of the many-body ground state, we follow its real-time evolution using the time-dependent density-matrix renormalization group. We observe that interactions lead to fractionalization of the initial excitation into counter-propagating left- and right-moving modes, whose energies can be resolved in real space. To interpret these results, we employ Luttinger liquid theory, which  allows us to derive analytical predictions for the energy carried by the emergent modes. We find good agreement between field-theoretical predictions and numerical simulations in the low-energy regime. In contrast to charge fractionalization, which  is completely determined by the Luttinger liquid parameter, we show that energy partitioning is non-universal and depends on details  of  the injected wave packet, such as its width. Our results provide a real-space characterization of energy partitioning  in one-dimensional systems and establish a quantitative comparison between non-equilibrium numerical simulations and the effective field-theory description.}}

\vspace{\baselineskip}

%%%%%%%%%% BLOCK: Copyright information
% This block will be filled during the proof stage, and finilized just before publication.
% It exists here only as a placeholder, and should not be modified by authors.
\noindent\textcolor{white!90!black}{%
\fbox{\parbox{0.975\linewidth}{%
\textcolor{white!40!black}{\begin{tabular}{lr}%
  \begin{minipage}{0.6\textwidth}%
    {\small Copyright attribution to authors. \newline
    This work is a submission to SciPost Physics. \newline
    License information to appear upon publication. \newline
    Publication information to appear upon publication.}
  \end{minipage} & \begin{minipage}{0.4\textwidth}
    {\small Received Date \newline Accepted Date \newline Published Date}%
  \end{minipage}
\end{tabular}}
}}
}

%\linenumbers

\vspace{10pt}
\noindent\rule{\textwidth}{1pt}
\tableofcontents
\noindent\rule{\textwidth}{1pt}
\vspace{10pt}

\section{Introduction}
\label{sec:intro}

 This study of spontaneously 
fractionalized energy transport in one-dimensional (1D) systems is
dedicated to the memory of Ian Affleck as part of the 
commemorative series.   
We welcome the opportunity to acknowledge his profound and lasting contributions to the theory of 1D quantum systems, which helped to shape and advance the field. Beyond his scientific achievements, he mentored and inspired many of us, 
thereby paving the path
to continue to develop cutting-edge work in this area. The present work builds upon Ian Affleck's numerous seminal contributions to the understanding of 1D quantum systems and collective quantum phenomena %\cite{Affleck1988} 
with notable works on spin chains\cite{aklt,Affleck1988,affleck1987,affleck1989,critical,affleck1992,susc,white1996}, %non-linear effects, 
quantum impurities in 1D \cite{kondo}, 
fractional ground states degeneracies \cite{g-degeneracy}, and 
fractional magnetization plateaus \cite{oshikawa1997} just to name a few.

Systems in 1D provide a natural setting for exotic many-body phenomena that have no direct counterpart in higher dimensions.   One of the most 
counter-intuitive effects is the natural occurrence of fractional charges, 
e.g.~in fractional quantum Hall edges \cite{fqhe}, 
quantum wires \cite{safi1995, pham2000, pham2002,LEHUR20083037,Steinberg2007,Kamata2014,Inoue2014,Perfetto2014,Calzona2015,Karzig2011,Calzona2016,Ramos2024},
and in the form of fractional edge spins in the famous
Affleck-Kennedy-Lieb-Tasaki (AKLT) model \cite{aklt}.  
In the latter work, Affleck and
co-workers predicted localized spin-1/2 edge states, which later served as 
the first physical realization where 1D fractionalization was observed experimentally
via electron spin resonance in spin-1 chains \cite{esr}.  

In this work we focus on 
gapless 
1D wires \cite{Deshpande2010,Giamarchi,Haldane1981}, 
where electron-like quasi-particles \cite{Landau1965} are
no longer 
 stable excitation of the interacting system. 
Instead, a directionally injected particle spontaneously partitions  
into independent right- and left-moving density waves that 
%propagate independently with the sound velocity and 
carry fractional 
portions of the injected charge, magnetization, and energy \cite{safi1995, pham2000, pham2002,LEHUR20083037,Steinberg2007,Kamata2014,Inoue2014,Perfetto2014,Calzona2015,Karzig2011,Calzona2016,Ramos2024}.
The fractionally transported charges are directly related to the 
Luttinger parameter and do not depend on the injection protocol.  
However, a different picture emerges 
for energy partitioning \cite{Karzig2011,Calzona2016}:  In an identical 
experiment the transported fractions of energy are different from 
the fractions of the charges. We now illustrate that quantitative predictions for non-universal energy partitioning 
of Gaussian wave packets
can obtained by a careful evaluation and integration of 
three-point functions.
%{\bf here we need to specify, what we do different compared to those two works \cite{Karzig2011,Calzona2016} maybe?}

Numerically, we obtain corresponding 
results for the energy partitioning in 1D wires using 
dynamic simulations, which show a strong dependence on the 
momentum and the width of wave packet in agreement with our
analytic results. 
More specifically, we create a Gaussian wave packet on top of the many-body ground state and follow its real-time evolution using the time-dependent density matrix renormalization group (tDMRG) \cite{Ramos2024,white2004}. In the presence of interactions, the injected excitation fractionalizes into counter-propagating left- and right-moving modes, whose individual energy contributions can be extracted from the dynamics. 
It should be noted that 
the previously predicted ''universal'' behavior of 
energy partition using point-like injection cannot be reached.

The paper is organized as follows. In Sec.~\ref{sec:model}, we introduce the model and the out-of-equilibrium protocol used to investigate energy partitioning. In Sec.~\ref{sec:numerical}, we present our numerical findings and highlight features that we aim to understand analytically using field theory.  In Sec.~\ref{sec:low-energy}, we develop the field-theoretical description and employ Luttinger liquid theory to derive predictions for the left- and right-moving energy contributions. Additionally, we systematically compare these predictions with our tDMRG results. Finally, we present our concluding remarks in Sec.~\ref{sec:conclusion}.

\section{Model and Out-of-Equilibrium Protocol}\label{sec:model}
To investigate the energy partitioning in a critical  1D system, we consider the following  spinless fermionic model
\begin{equation}
	H=\sum_{j=1}^{L-1}\left[-\frac{1}{2}\left(c^\dagger_jc^{\phantom\dagger}_{j+1}+\text{h.c.}\right)+Vn_jn_{j+1}\right],\label{eq:Ham}
\end{equation}
where $c_j$ ($c_j^\dagger$) annihilates (creates) a fermion at site $j$, $V$ denotes the nearest-neighbor interaction strength, $n_j=c_j^\dagger c^{\phantom\dagger}_j$ is the local density operator, and $L$ is total number of sites. The Hamiltonian has a global $U(1)$ symmetry, corresponding to the conservation of  the total particle number, $N=\sum_{j=1}^L n_j$. Throughout this work, we focus on the half-filled case, corresponding to a particle density of $n=N/L=1/2$. Through the Jordan-Wigner transformation, the Hamiltonian in Eq.~\eqref{eq:Ham} maps onto the spin-$1/2$ XXZ chain and is exactly solvable by means of the Bethe ansatz \cite{Bethe}. Its ground-state phase diagram consists of two gapped phases  for $|V|>1$, separated by a critical phase in the regime $-1<V\leq 1$, whose low-energy properties are described by LL theory \cite{Affleck1988}.

To probe the energy partitioning during the real-time dynamics, we prepare at $t=0$ a Gaussian wave packet on top of the ground state $|\Psi_0\rangle$ of $H$:
\begin{equation}
	|\Psi(t=0)\rangle=A\sum_{j=1}^{L}e^{-\frac{(j-j_0)^2}{2\sigma^2}}e^{ik_0 j}c^\dagger_{j}|\Psi_0\rangle,\label{eq:init}
\end{equation}
where $j_0$ is the wave-packet center, $\sigma^2/2$ its variance, $k_0$ the mean momentum, and $A$ a normalization constant. The state then evolves according to the time-dependent Schr\"odinger equation, $|\Psi(t)\rangle=e^{-iHt}|\Psi(t=0)\rangle$, where we set $\hbar=1$. The Hamiltonian has the form $H=\sum_{j=1}^{L-1}h_{j,j+1}$, where the Hamiltonian density $h_{j,j+1}$ is associated with each bond. We define a site-centered energy density operator $\mathcal{H}_j$ by taking the average between the two adjacent bonds containing each site $j$ in the bulk:\be
\mc H_j =\frac12\left(h_{j-1,j}+h_{j,j+1}\right).
\ee
In Ref. \cite{Ramos2024}, the excitation dynamics was monitored through the time-dependent local charge excess,\be
 n_j(t)=\langle \Psi(t)|n_j|\Psi(t)\rangle-\langle \Psi_0|n_j|\Psi_0\rangle.\label{chargenj}
\ee
In this work, we consider the time-dependent local energy  excess,
\begin{equation}
	\mathcal{E}_j(t)=\langle \Psi(t)|\mathcal{H}_j|\Psi(t)\rangle-\langle \Psi_0|\mathcal{H}_j|\Psi_0\rangle.\label{Ejsite}
\end{equation}
In the non-interacting limit $V=0$, the dynamics can be calculated exactly, providing a useful benchmark for our numerical simulations (see Appendix~\ref{app:free-fermions}).

\section{Numerical results \label{sec:numerical}}

In general, we investigate the effects of the wave-packet width and the mean momentum on the wave-packet dynamics using the tDMRG. In all numerical simulations hereafter, we focus on the regime of repulsive interactions $0 \le V \le 1$, and fix the system size to $L=300$ and  $j_0 = L/2$. To select low-energy excitations that can be described by LL theory, we choose $1\ll \sigma \ll L$ and $|k_0|\approx k_F$, where $k_F=\pi/2$ is the Fermi momentum at half filling.  In addition, the time evolution is numerically performed using the second-order Suzuki-Trotter (ST) decomposition with time step $\delta t=0.05$. Further details of the numerical simulations are provided in Appendix~\ref{app:est-acc}.

%By fine-tuning $k_0$ and $\sigma$, we can select the energy regime of interest. 

\begin{figure}
	\centering	\includegraphics[width=10cm]{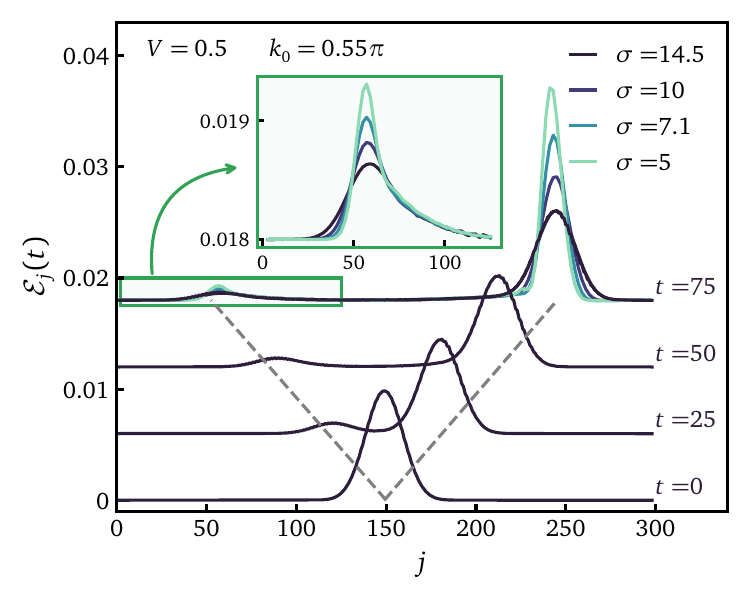}
	\caption{Snapshots of the energy excess for $V=0.5$, $k_0=0.55\pi$ and  the times indicated in the figure. The dashed lines indicate the light cone associated  to the propagation of the low-energy modes, defined by the velocity  $v\approx1.299$ calculated from Eq. (\ref{velocity}). For $t=75$, we show energy profiles for  $\sigma=14.5,\,10,\,7.1,$ and 5. The inset is a zoom-in of the left-moving hump at $t=75$.\label{fig:snp-sigma}}
\end{figure}

A representative numerical result of wave-packet dynamics is shown in Fig. \ref{fig:snp-sigma} for $V=0.5$ and distinct values of $\sigma$. This figure displays snapshots of the local energy excess at different times. We note that the two-bond average in Eq. (\ref{Ejsite})   smoothes out $2k_F$ oscillations in the energy density profile. We observe that the initially localized   excitation separates into left- and right-moving components, which propagate with the velocity  characteristic of the low-energy modes. The asymmetric amplitudes of the two wave packets already indicate that  the left- and right-moving excitations carry different energies.

\begin{figure}
	\centering\includegraphics[width=14cm]{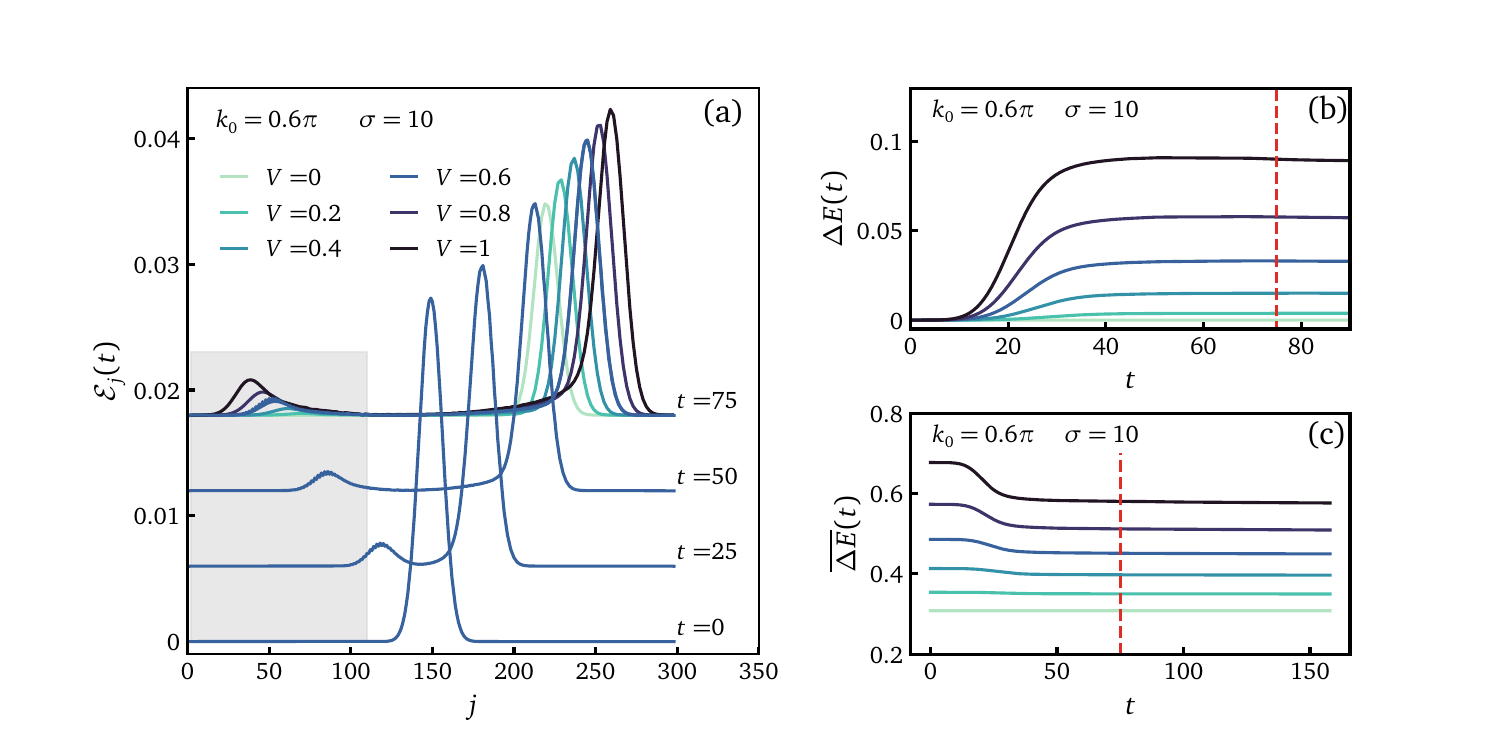}
	\caption{(a) Snapshots of the wave-packet dynamics for $k_0=0.6\pi$, $\sigma=10$ and several values of $V$. The shaded area indicates the left segment used to estimate $E_L$. Panels (b) and (c) respectively show the energies contained in the chain segments $j\in[1,j_\text{max}]$ and  $j\in[j_\text{max}+1,L-1]$, with $j_\text{max}=125$. The color coding for different values of $V$  in these panels correspond follows that in panel (a). The dashed lines indicate the time at which the estimates of $E_L$ and $E_R$ are extracted.\label{fig:snaps-app}}
\end{figure}

We denote by $E_L$ and $E_R$ the left- and right-moving energies. In practice, to estimate these quantities, we evolve the initial wave packet in time and define chain segments over which $\mathcal{E}_j(t)$ is integrated. In Fig.~\ref{fig:snaps-app}(a), we show snapshots of the energy profile for $k_0=0.6\pi$,  $\sigma=10$ and several values of $V$.  The time-dependent energies contained in the interval $j\in[1,j_\text{max}]$ and $j\in[j_\text{max}+1,L-1]$ are  respectively defined as 
\begin{eqnarray}
	\Delta E(t)&=&\sum_{j=1}^{j_\text{max}}\mathcal{E}_j(t),\\
	\overline{\Delta E}(t)&=&\sum_{j=j_\text{max}+1}^{L-1}\mathcal{E}_j(t).
\end{eqnarray}
The value of $j_{\rm max}$ is chosen such that the left and right segments can fully harbor the respective fractional wave packets, allowing us to extract their energies by analyzing the time-dependent energy contained in each segment. It is worth mentioning that there is some freedom in the choice of $j_{\rm max}$, provided that the segments are sufficiently large to contain the two wave packets separately and sufficiently narrow to delay the onset of contributions from revival signals. In particular, here, we set $j_{\rm max}=125$. Overall, at times when the left- and right-moving wave packets are well separated and fully contained within the complementary segments, the corresponding energies approach well-defined plateaus; see Figs.~\ref{fig:snaps-app}(b) and \ref{fig:snaps-app}(c). These plateau values are then taken as reliable estimates of the corresponding energies in the infinite-time limit. For instance, from Figs.~\ref{fig:snaps-app}(b) and \ref{fig:snaps-app}(c) we  estimate $E_L=\Delta E(t=75)$ and $E_R=\overline{\Delta E}(t=75)$.

Similarly, we denote by $Q_R$ and $Q_L$ the fractional charges carried by right and left movers,  respectively. We estimate these charges numerically using the same procedure described for the energy, replacing the energy density $\mc E_j$ by the charge density $n_j$. LL theory predicts that $Q_{R,L}$ only depend on the Luttinger parameter $K$, and are given by \cite{pham2000}\be
Q_{R,L}=\frac{1\pm K}{2}. \label{QRL}
\ee
For the model in Eq.~\eqref{eq:Ham} at half filling, the Luttinger parameter is known   analytically \cite{Giamarchi}:
\begin{eqnarray}
	K&=&\frac{\pi}{2\left[\pi - \arccos(V)\right]}.
\end{eqnarray} 
Note that   $Q_L$ vanishes at the non-interacting point, $K=1$, since there is no fractionalization of the wave packet in this case.

\begin{figure}
\centering\includegraphics[width=10cm]{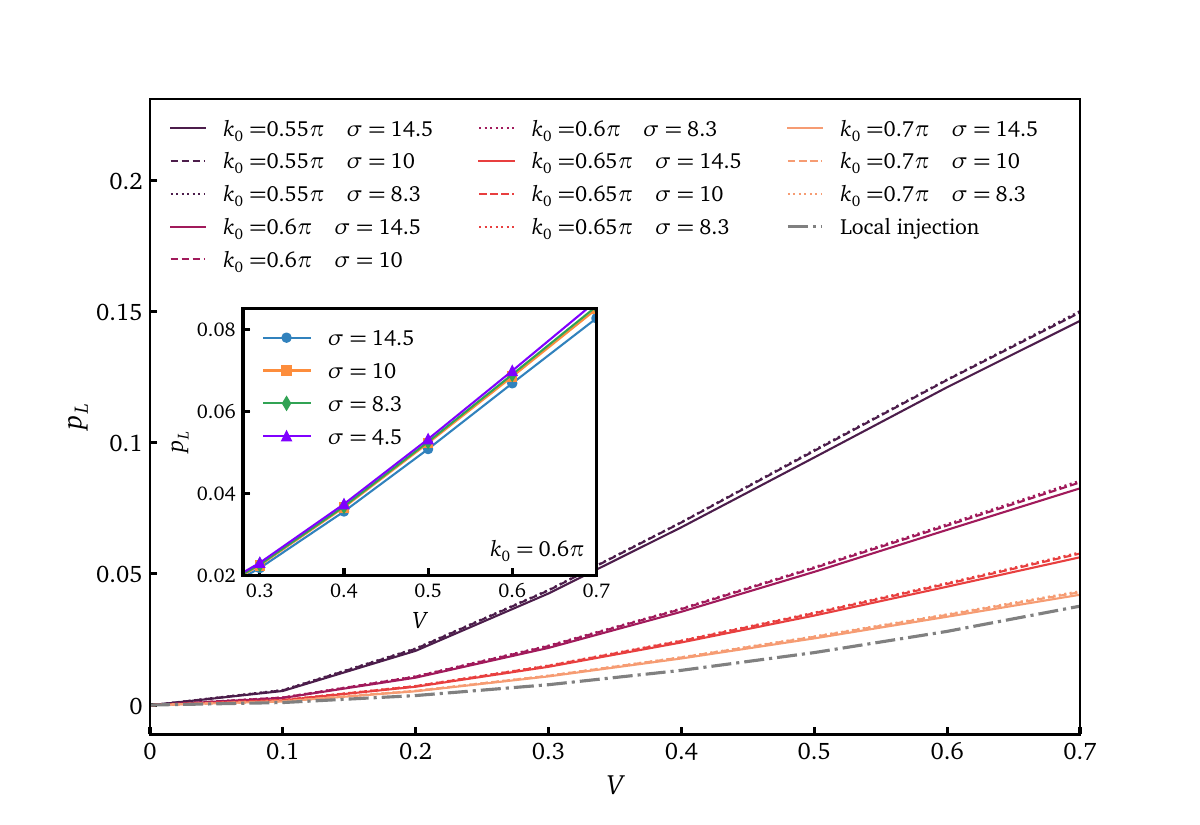}
	\caption{Fraction of the total energy propagating to the left as a function of the interaction strength $V$ for several values of $k_0$ and $\sigma$. The solid, dashed, and dotted lines correspond to $\sigma =$ 14.5, 10, and 8.3, respectively, with each color set denoting a fixed value of $k_0$. The dash-dotted line shows the prediction for local injection from Refs.~\cite{Karzig2011,Calzona2016}; see Eq.~(\ref{localinj}). The inset shows the results for $k_0 = 0.6 \pi$ and different values of $\sigma$ in the intermediate- to strong-interaction regime.   \label{fig:pl}}
\end{figure}

To characterize the   behavior of $E_{L,R}$, we define $p_{L}$ and $p_R$ as the fraction of the total energy that propagates to the left and to the right, respectively, so that  
\begin{equation}
	p_{L,R}=\frac{E_{L,R}}{E_L+E_R}.\label{eq:fraction}
\end{equation}
Figure \ref{fig:pl} shows $p_L$ as a function of the interaction strength $V$ for several values of $k_0$ and $\sigma$. Interestingly, the results clearly show that   $p_L$ decreases with increasing $k_0$. This behavior highlights the nontrivial momentum dependence of the excitation energy. In addition,  for  fixed values of $k_0$, our DMRG results show that $p_L$ increases as the wave-packet width decreases; see  the inset of Fig.~\ref{fig:pl}. 

It is known that the energy flow resulting from particle injection into a Luttinger liquid depends on the protocol \cite{Karzig2011,Calzona2016}. For the protocol studied in Refs.~\cite{Karzig2011,Calzona2016}, 
 the energy partitioning simplifies in the local-injection limit, where it depends only on the Luttinger parameter $K$, which  encodes the strength of interactions in the system.  The predicted result is 
\begin{equation}
	p^{\text{local-inj}}_{L,R}=\frac{(1\mp K)^2}{2(1+K^2)}.\label{localinj}
\end{equation} 
The prediction $p^{\text{local-inj}}_{L}$ is shown as a dash-dotted line in Fig. \ref{fig:pl}. In our protocol, the local injection in real space is obtained in the limit $\sigma\rightarrow0$. Decreasing $\sigma$, however, we find that  the numerical results move away from the predicted  limit.

\begin{figure}
\centering\includegraphics[width=8cm]{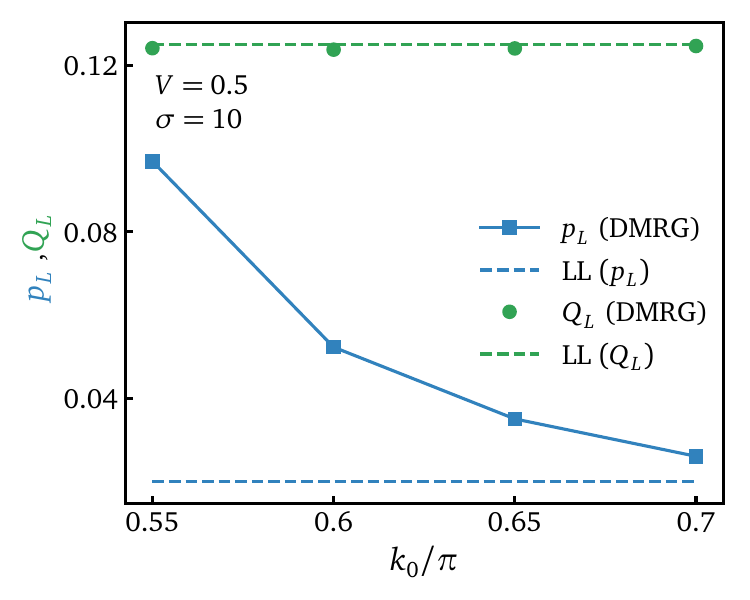}
\caption{$Q_L$ and $p_L$ as a function of $k_0$ obtained for $V=0.5$ and $\sigma=10$ . The symbols represent the DMRG estimates. The blue and green dashed lines are the LL predictions for $p_L$ and $Q_L$, respectively; see Eqs.~\eqref{QRL} and ~\eqref{localinj}.\label{fig:qlpl}}
\end{figure}

To illustrate the difference between energy and charge, in Fig.~\ref{fig:qlpl} we show a typical result for $Q_L$ and $p_L$ as a function of the mean momentum $k_0$ of the Gaussian wave packet, obtained for $V=0.5$ and $\sigma=10$. Our numerical results confirm that, at half filling, the fractional charges do not depend on $k_0$ and are in agreement with the LL prediction in Eq.~(\ref{QRL}) \cite{Ramos2024}. By contrast, the energy varies with $k_0$ and appears to approach the LL prediction as $k_0$ increases. However, this is not the appropriate low-energy limit, since increasing $k_0$ progressively shifts the wave packet toward higher-energy contributions, where the LL description is no longer expected to be valid.

Motivated by the features discussed here, in the following we provide a detailed field-theoretical interpretation of the problem and compare the predictions with our numerical findings.

\section{Low-Energy Description}\label{sec:low-energy}

The low-energy physics of the critical phase of model \eqref{eq:Ham}  is described by the LL theory, which relies on a linearization of the dispersion relation about the Fermi points \cite{Giamarchi}. Within this framework, the fermionic creation operator is expanded as
\begin{equation}
	c^\dagger_j\sim e^{\mathrm{i} k_F x}\Psi^\dagger_L(x)+ e^{-\mathrm{i} k_F x}\Psi^\dagger_R(x),
\end{equation}
where $\Psi_{L,R}(x)$ denote slowly varying fermionic fields associated with excitations with momentum near  $\mp k_F$. Upon bosonization, the effective low-energy Hamiltonian takes the form
\begin{equation}
	H_{\mathrm{LL}}= \sum_{\gamma=L,R}v\int dxT_\gamma(x)=\sum_{\gamma=L,R}H_\gamma,
\end{equation}
where $v$ is the sound velocity and $H_{L,R}$ are the chiral Hamiltonians. The chiral energy-density operators $T_{L,R}(x)$ are given by 
\begin{equation}
	T_{L,R}(x)=\frac{1}{2}\left[\partial_x\varphi_{L,R}(x)\right]^2.
\end{equation}
Here, $\varphi_{L,R}(x)$ denote the right- and left-moving components of the bosonic field, which obey $[\partial_{x}\varphi_\gamma(x),\varphi_{\gamma'}(x')]=i\gamma\delta_{\gamma\gamma'}\delta(x-x')$, with the chirality index   $\gamma=R,L=\pm$.  For the integrable model in Eq. (\ref{eq:Ham}) at half filling, the sound velocity is known analytically \cite{Giamarchi}:
\begin{equation}
	v=\frac{\pi \sqrt{1-V^2}}{2\arccos(V)}.\label{velocity}
\end{equation}

To construct the wave-packet excitation in the continuum, we consider the regime
$k_0>k_F$ and   $k_0+k_F\gg \sigma^{-1}$. The latter  ensures that   the wave function of the wave packet in momentum space  has negligible overlap with the left Fermi point. Under this condition, the contribution of the left-moving field $\Psi_L^\dagger(x)$ can be neglected, and the excitation is governed by the right-moving branch. In the continuum limit of an infinite system, the initial state is therefore written as
 \begin{equation} 
	|\Psi\rangle= \int_{-\infty}^{\infty}dx\, e^{ik_0x} e^{-\frac{x^2}{2\sigma^2}} e^{-ik_Fx} \Psi_R^\dagger(x)|0\rangle,\label{continuum}
\end{equation} 
where $\left|0\right\rangle$ is the bosonic vacuum.  Within the bosonization framework, the right-moving fermionic field can be factorized as
\begin{equation} 
\Psi_R^\dagger(x) = V_L^\dagger(x)V_R^\dagger(x),
\end{equation}
where  $V_{L,R}(x)$ are chiral vertex operators: 
\begin{equation}
	V_\gamma(x) = e^{i\sqrt{\frac{2\pi}{K}}\,Q_\gamma\,\varphi_\gamma(x)}. 
\end{equation}	
The coefficients $Q_L$ and $Q_R$ correspond to the   fractional charges in Eq. (\ref{QRL}). The corresponding scaling  dimensions of the vertex operators are
 \begin{equation}
\alpha_{L,R}=\frac{Q_{L,R}^2}{2K}.
 \end{equation}
 
Our aim now is  to determine how the injected energy is partitioned between the two chiral sectors. To do so, we  must evaluate the expectation values
\begin{equation}
	E_{L,R}=\frac{\langle \Psi |H_{L,R}|\Psi\rangle}{\langle \Psi |\Psi\rangle},\label{eq:erl}
\end{equation}
which quantify the energies transported by the right- and left-moving
excitations. Note that, since $[H_{LL},H_{L,R}]=0$, these quantities are conserved throughout the time evolution. Therefore, it  suffices to evaluate them at $t=0$.

Let us first discuss the energy transported by the left movers. Expressing the fermionic fields in terms of the chiral vertex operators, we obtain from Eqs.     \eqref{continuum} and (\ref{eq:erl})
\begin{equation}
	E_L=\frac{v}{\mathcal N}\int dx\,dx_1dx_2\,
	e^{iq(x_2-x_1)}
	e^{-\frac{(x_1^2+x_2^2)}{2\sigma^2}}
	\langle V_L(x_1)T_L(x)V_L^\dagger(x_2)\rangle
	\langle V_R(x_1)V_R^\dagger(x_2)\rangle,
	\label{eq:EL}
\end{equation}
where $\mathcal{N}$ is the squared norm of the state and the expectation value is computed in the ground state $|0\rangle$. It is instructive to compare this with the analogous expression for the   charge operator: \be
N_L=\frac{1}{\mathcal N}\int dx\,dx_1dx_2\,
	e^{iq(x_2-x_1)}
	e^{-\frac{(x_1^2+x_2^2)}{2\sigma^2}}
	\langle V_L(x_1)q_L(x)V_L^\dagger(x_2)\rangle
	\langle V_R(x_1)V_R^\dagger(x_2)\rangle,
	\label{eq:QL}
\ee
where $q_L(x)=\sqrt{\frac{K}{2\pi}}\partial_x\varphi_L(x)$ is the charge density associated with left movers.   The operator relation  ${[\int dx \, q_L(x),V^\dagger_L(x)]=Q_LV^\dagger_L(x)}$ implies that  the chiral vertex operator carries a well-defined charge.  Thus, the remaining integrals in Eq. (\ref{eq:QL}) reduce to the squared norm of the state $|\Psi\rangle$ and simply cancel $\mc N$ in the denominator, and we immediately obtain $N_L=Q_L$. By contrast, there is no such operator relation for the energy, i.e., the  state created by the action of the chiral vertex operators on the ground state  is \emph{not} an eigenstate of  $H_L$. As a consequence, the calculation of $E_L$ from Eq. (\ref{eq:EL}) actually requires computing the three-point function of chiral vertex operators with $T_L(x)$, and the scaling dimensions of both $V_R(x)$ and $V_L(x)$ can affect the result for $E_L$. 

As  discussed in Appendix~\ref{app:llcalc}, the integrals in the expression for $E_L$ in Eq. (\ref{eq:EL}) exhibit ultraviolet (UV) divergences and need to be regularized  by introducing short-distance cutoff parameters $\eta$ and $\epsilon$.  Such UV divergences are usually a sign that some aspects of the  property being calculated may depend on microscopic details. We obtain (see Appendix~\ref{app:llcalc})
\begin{equation}
	E_L=\frac{v}{2\sigma}\frac{Q_L^2}{K}\frac{a \, \sigma^{2\alpha} f_0(K)+f_2(q\sigma,K)}{b \, \sigma^{2\alpha-1} g_0(K)+g_1(q\sigma,K) }.\label{eq:ELresult}
\end{equation}
Here, $q=k_0-k_F$,  $\alpha=\alpha_L+\alpha_R=\frac{1+K^2}{4K}$, and we define the following  functions of the Luttinger parameter $K$ and the    dimensionless parameter  $u=q\sigma$:
\begin{eqnarray}
f_0(K)&=&g_0(K)=\frac{\sqrt{\pi } \Gamma \left(\alpha\right)}{\Gamma\left(2\alpha_R\right)},\\
f_2(u,K)&=&\Gamma \left(-\alpha \right) \, _1F_1\left(-\alpha;\frac{1}{2};-u^2\right),\\
g_1(u,K)&=&2 u \Gamma \left(1-\alpha\right) \, _1F_1\left(1-\alpha;\frac{3}{2};-u^2\right).
\end{eqnarray}
 In addition, $a =\epsilon^{-2\alpha}>0$ and $b=\eta^{1-2\alpha} $ are non-universal coefficients that depend on the short-distance cutoffs. Note that $v/\sigma$ has dimensions of energy; however,  due to the non-universal terms, the dimensionless quantity $\sigma E_L/v$ is \emph{not} a scaling function of $u=q\sigma$.  Generically, $a$ and $b$ vary with  the interaction strength but are  independent of the wave-packet parameters $k_0$ and $\sigma$. We also note that the analytical result in Eq. (\ref{eq:ELresult}) relies on an expansion in the short-distance cutoffs, which is valid provided that the integrals are dominated by large-distance contributions, but breaks down for $q\gtrsim \epsilon^{-1}, \eta^{-1}$.

The calculation of the energy transported by the right movers is analogous. We obtain
\begin{equation}
 E_R=\frac{v}{2\sigma}\frac{Q_R^2}{K}\frac{\overline{a}\,  \sigma^{2\alpha} \overline{f}_0(K)+\overline{c}  \sigma^{2\alpha-1} \,q\sigma \overline{f}_1(K) +\overline{f}_2(q\sigma,K)}{b \sigma^{2\alpha-1} g_0(K)+g_1(q\sigma,K) },\label{eq:ERresult}
\end{equation}
with 
\begin{eqnarray}
	\overline{f}_0(K)&=&\frac{\pi  2^{-2\alpha_L} \Gamma \left(2\alpha\right)}{\Gamma \left(2\alpha_L\right) \Gamma \left(\alpha+\frac{3}2\right)},\\
	\overline{f}_1(K)&=&\frac{\sqrt{\pi }  \Gamma \left(\alpha\right)}{\Gamma \left(\alpha+\frac{3}2\right)},\\
	\overline{f}_2(u,K)&=&-f_2(u,K)=-\Gamma \left(-\alpha\right) \, _1F_1\left(-\alpha;\frac{1}{2};-u^2\right).
\end{eqnarray}
The coefficient $b$ in Eq.~\eqref{eq:ERresult} is the same as that appearing in Eq.~\eqref{eq:ELresult} and is fixed by the normalization factor $\mathcal{N}$. The coefficients $\overline{a}$ and $\overline{c}$ are likewise non-universal and independent of $k_0$ and $\sigma$. In contrast to $E_L$, the numerator of $E_R$ contains an additional non-universal contribution with a linear dependence in  $u=q\sigma$.

A useful consistency check is provided by the non-interacting limit, in which  $K\rightarrow1$ and $v\to1$. In this regime, the injected particle remains a free excitation, and the entire energy is carried by the right-moving wave packet. Taking this limit in the analytical expression, we obtain
\begin{eqnarray}
	\overline{f}_0(K\rightarrow1)&=&0,\\
	g_0(K\rightarrow1)&=&\frac{1}{2}\overline{f}_1(K\rightarrow1)
	=\frac{1}{\operatorname{erf}(u)}g_1(u,K\rightarrow1)=\pi,\\
	\overline{f}_2(u,K\rightarrow1)&=&2\sqrt{\pi}e^{-u^2}+2\pi u\operatorname{erf}(u),
\end{eqnarray}
where $\operatorname{erf}(x)$ denotes the error function. Consequently, Eq.~\eqref{eq:ERresult} reduces to
\begin{equation}
	E_R\rightarrow\frac{1}{\sigma}
	\frac{e^{-u^2}+\sqrt{\pi}u\left[\overline{c}+\operatorname{erf}(u)\right]}
	{\sqrt{\pi}\left[b+\operatorname{erf}(u)\right]}.
\end{equation}
This expression agrees with the direct analytical evaluation of the energy carried by a Gaussian wave packet with a linear dispersion relation measured relative to the Fermi momentum,
\begin{equation}
	E_R=\frac{\int_0^\infty dk k e^{-\sigma^2(k-q)^2}}
	{\int_0^\infty dk e^{-\sigma^2(k-q)^2}}.
\end{equation}
Therefore, the field-theoretical expression correctly reproduces the expected free-fermion limit.

It is worth mentioning that it is possible to construct eigenstates of the chiral Hamiltonians $H_{L,R}$ from the chiral vertex operators \cite{pham2000}.  These eigenstates are obtained by taking a Fourier transform $\int_0^\ell  dxe^{-iqx}V^\dagger_{\gamma}(x) |0\rangle$, with quantized  momenta $q=\frac{2\pi n}{\ell} $ in a finite system with length $\ell$ and periodic boundary conditions. The corresponding energies have the form $E_{\gamma,n}=\frac{\pi v}{2\ell}\frac{Q_\gamma^2}{K}+\frac{2\pi v |n|}{\ell}$, where the first, momentum-independent  term involves the scaling dimension of the chiral vertex operator, while the second term has a linear momentum dependence and is associated with the contribution from chiral phonons  \cite{pham2000}. However, the energies $E_{\gamma}$ given by Eqs. (\ref{eq:ELresult}) and (\ref{eq:ERresult}) cannot be obtained by simply averaging  $E_{\gamma,n}$ with the wave function in momentum space. Instead, the energy expectation value is determined by the full correlation-function calculation, which in the interacting case  involves a product of correlators of right and left movers.

%Let us now turn to the comparison with  our numerical results. The time evolution enables the spatial identification of the right- and left-moving excitations. Once they can be individually resolved, we integrate $\mathcal{E}_j(t)$ over appropriate chain segments to isolate each contribution. Following this protocol, we obtain estimates of $E_L$ and $E_R$. A detailed discussion of the systematic procedure used to extract these quantities is provided in Appendix~\ref{app:est-acc}. \textcolor{magenta}{This paragraph might change if we explain this procedure in Sec. 3.}

 In contrast to charge fractionalization, the energy partitioning is not universal. In fact, the interaction dependence of $E_{L,R}$ is contained not only in the Luttinger parameter, but also in the non-universal coefficients. Nevertheless, the LL theory predicts the dependence  on the wave-packet parameters $k_0$ and $\sigma$. To test this prediction, we perform, for each fixed value of $V$, a simultaneous fit of the corresponding datasets to Eqs.~\eqref{eq:ELresult} and \eqref{eq:ERresult}. The non-universal coefficients are treated as free parameters and determined through a regularized nonlinear least-squares procedure. 
 
 To analyze the results, we first note that the non-universal terms in Eqs.~\eqref{eq:ELresult} and \eqref{eq:ERresult} differ in their dependence on $\sigma$.  For the energy carried by the left movers, we define the rescaled quantity\bea
 	\mathcal{E}_{L}&=& q\frac{2\sigma K}{v Q_{L,R}^2}\, E_L \nonumber\\
	&=& \frac{a q^{1-2\alpha} \,u^{2\alpha}   f_0(K)+q f_2(u,K)}{b \, q^{1-2\alpha}u^{2\alpha-1} g_0(K)+g_1(u,K) }.\label{mcEL}
 \eea
Looking at the leading terms in the numerator and in the denominator, we see that  the explicit momentum dependence that breaks the scaling behavior as a function of $u=q\sigma$ is proportional to $q^{1-2\alpha}$.  For weak interactions, we have $K\approx 1-\frac{2V}{\pi}$ and  $\alpha\approx \frac12+     \mc O(V^2)$. Thus, the small exponent in $q^{1-2\alpha}$ suggests an approximate data collapse of  $\mc E_L$ as a function of $u$, for different values of $q$, in the weakly interacting regime. Similarly, we define the rescaled energy for right movers\bea
 	\mathcal{E}_{R}&=&\frac{2\sigma K}{v Q_{R}^2},\nonumber\\
	&=&\frac{\overline{a}\, q^{-2\alpha} u^{2\alpha} \overline{f}_0(K)+\overline{c} q^{1-2\alpha} u^{2\alpha}   \overline{f}_1(K) +\overline{f}_2(u,K)}{b \, q^{1-2\alpha}u^{2\alpha-1} g_0(K)+g_1(u,K) }.\label{mcER}
 \eea 
 In this case, the first term in the numerator is proportional to $q^{-2\alpha}$, which indicates a stronger   momentum dependence for $\alpha\approx \frac12$. However, the factor  $\bar f_0(K)\sim \mc O(V^2)$ strongly suppresses this term for $V\ll 1$. Thus, we also expect an approximate data collapse in $\mc E_R$ as a function of $u$ in the weakly interacting regime.

\begin{figure}
	\centering	\includegraphics[width=13.8cm]{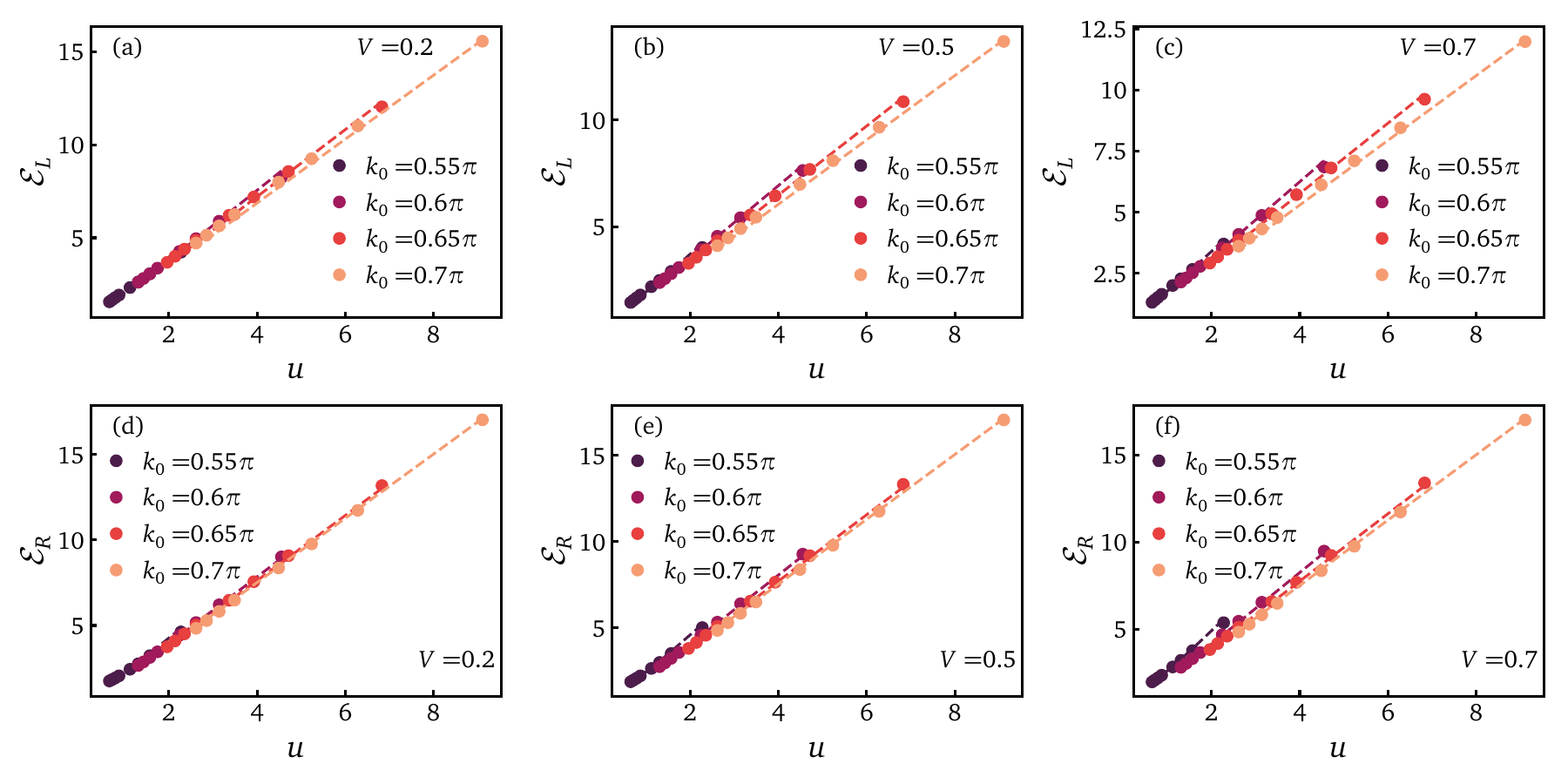}
	\caption{ Rescaled energy  transported by the (a)–(c) left-moving and (d)–(f) right-moving excitations as a function of $u$ for different values of $V$. Each dataset corresponds to a fixed value of $k_0$, as indicated in the legend. The symbols represent the DMRG results, while the dashed lines are fits to our data using Eqs.~\eqref{mcEL} and \eqref{mcER}. \label{fig:erl-dmrg}}
\end{figure}

In Fig.~\ref{fig:erl-dmrg}, we show the   numerical results for $\mc E_{L,R}$  as a function of $u$. The panel columns correspond to distinct values of $V$, while each curve is obtained for fixed $k_0$.
 For each panel, all dashed curves are obtained using the same set of fitting parameters. In our fitting procedure, we first fit $\mathcal{E}_L$ to determine the coefficients $a$ and $b$. Having obtained $b$, we then fit $\mathcal{E}_R$ and extract the remaining coefficients $\overline{a}$ and $\overline{c}$. Overall, we find excellent agreement between the LL predictions and the tDMRG results. Note, in particular, the data collapse in the weakly interacting case $V=0.2$.

Finally, let us comment on the discrepancy with the prediction for the local-injection limit in Eq.~(\ref{localinj}), 
which was derived previously by  taking $\sigma\to0$ for Luttinger Liquids with
purely linear dispersion relations \cite{Karzig2011,Calzona2016}. 
However, a narrow real-space width implies  a broad momentum-space distribution, with contributions spanning the entire energy spectrum, which is in general non-linear. 
In particular,
in any lattice model the time evolution of such a highly localized  single-particle excitation is influenced by modes with  energies outside  the linear-dispersion regime \cite{Pereira2009,Pereirareview}. Our  results indicate  that LL theory is applicable for  smooth density profiles, where the wave-packet width $\sigma$ is larger than the short-distance cutoffs, which are of the order of the lattice spacing. In this regime, however, the non-universal terms give      significant contributions     to the energies $E_{L,R}$, and the prediction for the  local-injection limit  is not recovered.

\section{Conclusion}\label{sec:conclusion}

We have studied the non-equilibrium dynamics of wave packets in a critical fermionic chain, focusing on the real-space partitioning of injected energy. Using tDMRG, we have shown that an initially localized excitation fractionalizes into counter-propagating left- and right-moving wave packets, whose individual energy contributions can be resolved during the time evolution. Within the low-energy regime, Luttinger liquid theory provides a description of this process and yields accurate predictions for the energy carried by the emergent modes. The comparison with numerical results demonstrates good agreement in the regime where the linearized description applies. Importantly, our results demonstrate that, in contrast to charge fractionalization, energy partitioning is generally non-universal and depends on the details of the injected wave packet. This sensitivity reflects the role of the excitation protocol in determining the distribution of energy among fractional modes. Finally, our work establishes a direct connection between non-equilibrium real-space dynamics and field-theoretical predictions, providing a complementary route to characterizing fractionalization phenomena beyond equilibrium properties.

\section*{Acknowledgements}

This work was supported by the Deutsche Forschungsgemeinschaft (DFG, German Research Foundation)-Project No. 277625399-TRR 185 OSCAR (A4,A5). R. G. P. acknowledges funding from Finep (Grant No. 1699/24 IIFFINEP) and  the Conselho Nacional de Desenvolvimento Cient\'ifico e Tecnol\'ogico (CNPq) (Grants No. 309569/2022-2 and 404274/2023-4). F. B. R. acknowledges funding from CNPq (Grant Project No. 446378/2024-0). The authors thank the high-performance clusters Elwetritsch and NPAD for providing computational resources.

\begin{appendix}
\numberwithin{equation}{section}

\section{Free fermions: $V=0$}\label{app:free-fermions}
The energy excess in the non-interacting limit $V=0$ can be calculated exactly. In this case, the initial state can be expressed in terms of single-particle eigenstates of $H$. For open boundary conditions, we find that $\mathcal{E}_j(t)$ is given by
\begin{equation}
	\mathcal{E}_j(t)=-\text{Re}\left[\sum_{k>kF}\sin(kj)e^{-i\cos(k) t }q^*_k\sum_{k>kF}\sin[k(j+1)]e^{i \cos(k) t }q_k\right],\label{eq:ff-lattice}
\end{equation}
where $q_k=\frac{2}{L+1}\sum_{j=1}^L f_j \sin(k j)$ and $f_j=A e^{ik_0j} e^{-\frac{(j-j_0)^2}{2\sigma^2}}$. 

\begin{figure}
\centering	\includegraphics[width=9cm]{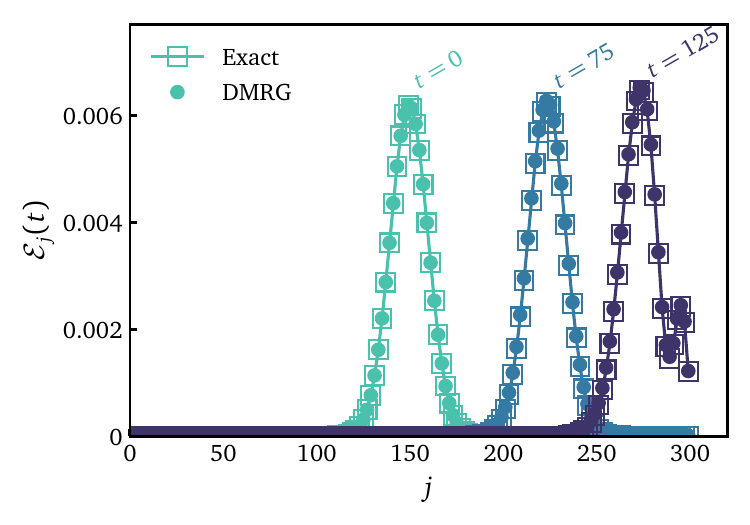}
\caption{Snapshots of the local energy excess $\mathcal{E}_j(t)$ for $V=0$, $k_0=0.55\pi$, and $\sigma=14.5$. The time corresponding to each snapshot is indicated above the wave packet. Filled circles show the tDMRG results and squares denote the exact results obtained from Eq. \eqref{eq:ff-lattice}. \label{fig:comparison-ff}}
\end{figure}

In Fig.~\ref{fig:comparison-ff}, we compare our tDMRG results with the exact solution obtained from Eq. \eqref{eq:ff-lattice}. We find excellent agreement between the numerical and analytical results across all times and positions.

\section{Left Energy : $E_L$}\label{app:llcalc}

Here, we provide a detailed derivation of Eq.~\eqref{eq:ELresult}. We start by calculating the squared norm $\mathcal{N}=\langle\Psi|\Psi\rangle$ for the state in the continuum limit given in Eq. (\ref{continuum}). Using the   two-point correlation functions for the chiral vertex operators, we obtain
\begin{equation}
	\mathcal{N}=\int dx_1dx_2\,	e^{iq(x_2-x_1)}	e^{-\frac{x_1^2+x_2^2}{2\sigma^2}}\frac{1}{[\eta+i(x_1-x_2)]^{2\alpha_L}}	\frac{1}{[\eta-i(x_1-x_2)]^{2\alpha_R}},
\end{equation}
where $q=k_0-k_F$ and $\eta>0$ is the short-distance cutoff associated with the time ordering of the vertex operators. Introducing the center-of-mass and relative coordinates,
\[
X=\frac{x_1+x_2}{2},\qquad
r=x_1-x_2,
\]
and integrating over $X$, we obtain 
\begin{equation}
	\mathcal N=	\sqrt{\pi}\,	2^{1-2\alpha} q^{2\alpha -2}	u^{2-2\alpha}	g(u;\alpha_L,\alpha_R),\label{eq:norm}
\end{equation}
with $u=q\sigma$ and $\alpha=\alpha_R+\alpha_L=\frac{1+K^2}{4K}$. The dimensionless function
\begin{equation}
	g(u;\alpha_L,\alpha_R) = \int_{-\infty}^{\infty} dy\,\frac{e^{-2iuy-y^2}}{(\tilde\eta+iy)^{2\alpha_L}
		(\tilde\eta-iy)^{2\alpha_R}},\label{eq:gintegral}
\end{equation}
depends on the scaling dimensions of the vertex operators and the dimensionless cutoff $\tilde\eta=\eta/\sigma$.
 Note that we have $\frac12\leq\alpha<\frac58$ for $\frac12<K\leq1$. For reference, in the non-interacting case $K=1$,   where $\alpha_R=1/2$ and $\alpha_L=0$, consider $u=0$. In this case, the integral becomes
\begin{equation}
g(0;0,1/2)=\int_{-\infty}^{\infty}d y\, \frac{e^{- y^2}}{ \tilde \eta -i   y}=\pi e^{\tilde \eta^2}\text{erf}(\tilde \eta)\xrightarrow{\tilde \eta\to0}\pi,
\end{equation}
which shows that the state has a finite norm in the non-interacting case.  However, for arbitrarily weak interaction, we have $2\alpha>1$ and  the integral  diverges if we take the limit $\tilde \eta\to0$. Extracting the leading singularity from the integral in Eq. (\ref{eq:gintegral}), we obtain 
\begin{eqnarray}
g(u;\alpha_L,\alpha_R)&=&\frac{\pi \Gamma(2\alpha-1)}{2^{2\alpha-2}\Gamma(2\alpha_L)\Gamma(2\alpha_R)}\, \tilde\eta^{1-2\alpha}+\cos [\pi  (\alpha_R-\alpha_L)] \Gamma \left(\frac{1}2-\alpha\right) \, _1F_1\left(\frac{1}{2}-\alpha;\frac{1}{2};-u^2\right)\nonumber\\
&&+2 u \sin [\pi (\alpha_R-\alpha_L)] \Gamma (1-\alpha) \, _1F_1\left(1-\alpha;\frac{3}{2};-u^2\right),
\end{eqnarray}
where $\Gamma(x)$ is the Gamma function and $_1F_1\left(a;b;z\right)$ is the confluent hypergeometric function of the first kind.

We next evaluate the numerator of Eq.~(\ref{eq:EL}), which is determined by the three-point correlation function $\langle V_L(x_1)T_L(x)V_L^\dagger(x_2)	\rangle$. The expansion of the vertex operators in powers of the bosonic field gives
\begin{eqnarray}
	\langle V_L(x_1)T_L(x)V_L^\dagger(x_2)\rangle &=& \frac12 \sum_{m,n=0}^{\infty}\frac{(-i\sqrt{4\pi\alpha_L})^m}{m!}\frac{(i\sqrt{4\pi\alpha_L})^n}{n!}\nonumber\\
	&&\times\left\langle:\varphi_L^m(x_1):	:[\partial_x\varphi_L(x)]^2::\varphi_L^n(x_2):\right\rangle.
\end{eqnarray}

Now, applying Wick's theorem to evaluate all possible contractions and introducing an imaginary-time splitting, $\delta\tau=\epsilon/(2v)$, yields
\begin{eqnarray}
\langle	V_L(x_1,\delta\tau)	T_L(x,0)V_L^\dagger(x_2,-\delta\tau)\rangle	&=&	\frac{\alpha_L}{2\pi[\epsilon+i(x_1-x_2)]^{2\alpha_L}}	\nonumber\\
&&\times\left[\frac{1}{\epsilon/2+i(x_1-x)}	+\frac{1}{\epsilon/2+i(x-x_2)}\right]^2,\label{eq:3point}
\end{eqnarray}
where we have used the bosonic correlation
\begin{equation}
	\langle	\varphi_L(x_1,\eta)\partial_x\varphi_L(x,0)\rangle	=\frac{i}{2\pi}	\frac{1}{\eta+i(x_1-x)}.
\end{equation}
Substituting Eq.~\eqref{eq:3point} into Eq.~(\ref{eq:EL}), introducing the coordinates $X$ and $r$, and performing the integration over $X$, we obtain 
\begin{eqnarray}
	E_L=\frac{2^{1-2\alpha}\sqrt{\pi}v\alpha_L q^{2\alpha -1}u^{1-2\alpha}}{\mc N} f(u; \alpha_L,\alpha_R),\label{eq:ELratio}
\end{eqnarray}
with 
\begin{equation}
f(u;\alpha_L,\alpha_R)=   \int_{-\infty}^{\infty}  dy\, \frac{e^{-i2u  y- y^2}}{( \tilde \epsilon+i y)^{2\alpha_L+1}( \tilde \epsilon-i y)^{2\alpha_R}}  ,\label{eq:fintegral}
\end{equation}
and $\tilde \epsilon=\epsilon/\sigma$. As before, this integral is UV divergent for $\tilde \epsilon\to 0$. Similar to $g(u;\alpha_L,\alpha_R)$, we extract the leading singularity so that
\begin{eqnarray}
f(u;\alpha_L,\alpha_R)&=& \frac{ \pi  \Gamma (2 \alpha)}{2^{2 \alpha-1} \Gamma (2\alpha_L+1) \Gamma (2\alpha_R)}\,\tilde\epsilon ^{-2 \alpha}-\frac{ \pi  [2 (\alpha_L-\alpha_R)+1] \Gamma (2 \alpha-1)}{2^{2 \alpha-2} \Gamma (2\alpha_L+1) \Gamma (2\alpha_R)}\tilde \epsilon ^{1-2 \alpha}\, u\nonumber\\
&&-2 u \cos [\pi  (\alpha_R-\alpha_L)] \Gamma \left(\frac{1}{2}-\alpha\right) \, _1F_1\left(\frac{1}{2}-\alpha;\frac{3}{2};-u^2\right)\nonumber\\
&&+\sin [\pi  (\alpha_R-\alpha_L)] \Gamma (-\alpha) \, _1F_1\left(-\alpha;\frac{1}{2};-u^2\right).
\end{eqnarray}
Combining this result with Eq. \eqref{eq:norm}, the left-moving energy takes the form
\begin{equation}
E_L=\frac{ v\alpha_L}{\sigma}\frac{f(u;\alpha_L,\alpha_R)}{f(u;\alpha_L,\alpha_R)}. \label{eq:EL2}
\end{equation}

For later convenience, we write the functions $f(u;\alpha_L,\alpha_R)$ and $g(u;\alpha_L,\alpha_R)$ as
\begin{eqnarray}
f(u;\alpha_L,\alpha_R)&=&f_0\tilde \epsilon^{-2\alpha}+f_1\tilde\epsilon^{1-2\alpha}+f_2,\\
g(u;\alpha_L,\alpha_R)&=&g_0\tilde\eta^{1-2\alpha}+g_1,
\end{eqnarray}
where, in terms of the scaling dimensions, the coefficients f's and g's are 
\begin{eqnarray}
	f_0(K)&=&g_0(K)=\frac{\sqrt{\pi } \Gamma \left(\alpha\right)}{\Gamma\left(2\alpha_R\right)},\\
	f_2(u,K)&=&\Gamma \left(-\alpha \right) \, _1F_1\left(-\alpha;\frac{1}{2};-u^2\right),\\
	g_1(u,K)&=&2 u \Gamma \left(1-\alpha\right) \, _1F_1\left(1-\alpha;\frac{3}{2};-u^2\right),
\end{eqnarray}

Finally, restoring the $\sigma$ dependence in the cutoffs and using $Q_L=(1-K)/2$, we find   
\begin{equation}
E_L=\frac{v(K)}{\sigma}\frac{Q_L^2}{2K}\frac{a \,q^{-2\alpha}u^{2\alpha} f_0(K)+f_2(u,K)}{b \,q^{1-2\alpha}u^{2\alpha-1} g_0(K)+g_1(u,K) },\label{eq:ELresultapp}
\end{equation}
where $a=\epsilon^{-2\alpha}>0$ and $b=\eta^{1-2\alpha}>0$ are non-universal prefactors that vary with the interaction.

Finally, we can apply Eq. \eqref{eq:ELresultapp} as long as the integrals in Eq.~\eqref{eq:gintegral} and Eq.~\eqref{eq:fintegral} are dominated by large-distance contributions.  As we increase $u$, the oscillating factor $e^{-i2u y}$ suppresses the large-distance contributions, and the analytical expressions deviate from the full result for the integrals when $q\gtrsim  \text{max}\{\eta^{-1},\epsilon^{-1}\} $.  In practice, to reach $u=q\sigma\gg1$  in the numerics requires entering the nonlinear dispersion regime, where the Luttinger liquid description is no longer reliable.
\section{ Accuracy of the Time Evolution }\label{app:est-acc}

 The source of error associated to our tDMRG results has two origins: the truncation error,  which is controlled by the number of kept states, and the Trotter error, which originates from  the ST decomposition implemented for the time evolution. For the former, we use up to 400 states in the truncated Hilbert space, for which the largest discarded error in the sweeps is smaller than $10^{-10}$. On the other hand, the Trotter error of a decomposition of order $n$ is $\sim(\delta t)^{n+1}$. As mentioned in the main text, we set the Trotter step to $\delta t=0.05$ in our simulations.

\begin{figure}
	\centering\includegraphics[width=9cm]{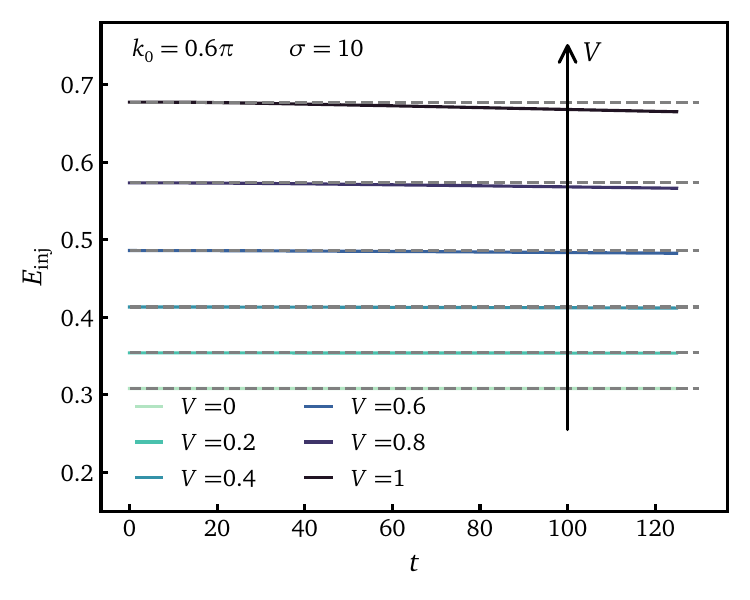}
	\caption{Total injected energy as a function of time for $k_0=0.6\pi$, and $\sigma=10$ and different values of $V$. The dashed lines, obtained at $t=0$, serve as benchmark for the errors introduced by the ST decomposition.\label{fig:Trotter-err} }
\end{figure}

 To assess the accuracy of the time evolution, we monitor the conservation of the injected energy. This quantity is given by the sum of  $\Delta E(t)$ and $\overline{\Delta E}(t)$, i.e. $E_{\mathrm{inj}} = \Delta E(t) + \overline{\Delta E}(t)$.
 Figure~\ref{fig:Trotter-err} displays a typical result of $E_{\mathrm{inj}}$ as a function of time. For comparison, we also include the injected energy computed at $t=0$ as a reference value, allowing us to quantify deviations that arise during the time evolution. For fixed $\delta t$, the simulation accuracy decreases with increasing  $V$. This trend stems from the Suzuki-Trotter decomposition, which expands the time-evolution operator in terms of nested commutators between the even- and odd-bond Hamiltonian components. Because the contributions of these commutators depend explicitly on  both $V$ and $\delta t$, numerical accuracy can be systematically improved by employing higher-order Suzuki-Trotter decompositions or reducing the time step.

\end{appendix}

\bibliography{references.bib}

\end{document}